\documentclass[prd,aps,twocolumn,superscriptaddress,floatfix,nofootinbib,eqsecnum]{revtex4-2}

\pdfoutput=1

\usepackage{amsmath}
\usepackage{amssymb}
\usepackage{bm}
\usepackage{graphicx}
\usepackage{dcolumn}
\usepackage[dvipsnames]{xcolor}
\usepackage{orcidlink}
\usepackage{hyperref}
\hypersetup{colorlinks=true,linkcolor=blue,citecolor=blue,urlcolor=blue}

\makeatletter
\def\frontmatter@affiliationfont{\normalfont\selectfont}
\makeatother
\DeclareRobustCommand{\emph}[1]{{\upshape #1}}

\newcommand{\FDM}{\mathcal{F}_{\rm DM}}
\newcommand{\Msun}{M_\odot}
\newcommand{\NM}{{\rm NM}}
\newcommand{\DM}{{\rm DM}}
\newcommand{\gcm}{\mathrm{g\,cm^{-3}}}

\begin{document}

\title{Impact of a fermionic dark-matter core on the structure, tidal deformability,\\ and moment of inertia of neutron stars}

\author{A.~R.~Sarma\,\orcidlink{0009-0005-5344-1189} }\email{abhiramisarmaa@gmail.com}
\affiliation{Department of Physics, Sardar Vallabhbhai National Institute of Technology, Surat 395007, Gujarat, India}

\author{A.~C.~Khunt\,\orcidlink{0000-0002-5615-4645}}\email{ack.gravity@gmail.com}
\affiliation{Department of Physics, Sardar Patel University, Vallabh Vidyanagar 388120, Gujarat, India}

\author{V.~K.~Ojha\,\orcidlink{0000-0002-0641-4015}}\email{vko@phy.svnit.ac.in}
\affiliation{Department of Physics, Sardar Vallabhbhai National Institute of Technology, Surat 395007, Gujarat, India}

\begin{abstract}
Dark matter (DM) accumulated inside neutron stars can alter their observable properties. We model such stars as two perfect fluids coupled only through gravity: $\beta$-equilibrated nucleonic matter described by the HPU-3 relativistic mean-field equation of state, and self-interacting fermionic DM described by a Walecka-type model. For DM mass fractions $\FDM=0$--$10\%$ we compute mass--radius sequences, tidal deformabilities and moments of inertia. In all stars with $M\geq0.6\,\Msun$ the DM forms a core that makes the star more compact. Raising $\FDM$ to $10\%$ lowers the maximum mass from $2.351$ to $2.115\,\Msun$ and shrinks a $1.4\,\Msun$ star from $13.44$ to $12.64$~km, while $\Lambda_{1.4}$ falls from $728$ to $461$. HPU-3 alone is too stiff for GW170817 ($\tilde\Lambda\simeq840$--$850$ against the 90\% bound $\tilde\Lambda\leq720$), but $\FDM\gtrsim3\%$ restores agreement while the maximum mass stays above $2.1\,\Msun$, consistent with PSR~J0348+0432 and PSR~J0740+6620. Tidal data alone therefore cannot distinguish a stiff nucleonic star with a few percent of DM from a softer star without DM. For the softer SLy EoS the heavy pulsars limit $\FDM$ to about $1.7\%$. The results are robust across the $1\sigma$ range of the DM parameters; DM particles lighter than about $0.5$~GeV instead form a halo. For co-rotating fluids the I--Love relation remains universal to better than $1\%$, whereas the compactness--Love relation shifts by up to $2.5\%$.
\end{abstract}


\maketitle

\section{Introduction}\label{sec:intro}

Neutron stars (NSs) let us study matter at densities, pressures and gravitational fields that no laboratory can reach~\cite{lattimer2004,lattimer2012annrev}. Their bulk properties are set by the equation of state (EoS) of dense matter, that is, the relation between pressure and energy density, and above nuclear saturation density this relation is still poorly known~\cite{oertel2017,lattimer2016physrep}. In the innermost core new particles such as hyperons, meson condensates or deconfined quarks may appear~\cite{lattimer2004,Lattimer2006}. Each of them softens or stiffens the EoS and so changes what we can observe, including the mass--radius ($M$--$R$) relation and the tidal deformability~\cite{Lattimer2006,Yunes2022}.

In general relativity (GR) a static, spherically symmetric NS is fixed by the equations of hydrostatic equilibrium, and their solutions depend strongly on the EoS. Different models of dense matter therefore predict different masses, radii, compactnesses and tidal responses, which is exactly why these quantities are useful probes of the interior.

Mass and radius measurements constrain the EoS most directly. Shapiro-delay timing of PSR~J1614$-$2230~\cite{demorest2010}, PSR~J0348+0432~\cite{antoniadis2013} and PSR~J0740+6620~\cite{Fonseca2021} showed that stars of $\approx2\,\Msun$ exist, which rules out many soft EoSs. The black-widow pulsar PSR~J0952$-$0607, with $M=2.35\pm0.17\,\Msun$~\cite{romani2022}, may be heavier still. Radii have come from the Neutron star Interior Composition Explorer (NICER), which models the pulse profiles of thermal X-ray emission from hot spots on the stellar surface. NICER results are available for the isolated pulsar PSR~J0030+0451~\cite{riley2019,miller2019nicer}, the massive PSR~J0740+6620~\cite{Riley2021,Miller:2021qha} and the nearby PSR~J0437$-$4715~\cite{Choudhury:2024xbk}, and Ref.~\cite{Rutherford:2024srk} discusses what they imply together for the EoS. Gravitational waves add a different kind of information. The binary NS merger GW170817~\cite{abbott2017gw170817} gave the first measurement of the tidal deformability $\Lambda$, which describes how strongly a star develops a quadrupole moment in the tidal field of its companion~\cite{Abbott2018,Abbott2019PRX}. Since $\Lambda$ depends on both the compactness and the density profile inside the star, it complements mass and radius measurements.

NSs are also a natural place to look for dark matter (DM). Evidence from galaxy dynamics and gravitational lensing~\cite{clowe2006} and from the cosmic microwave background~\cite{planck2018} shows that DM makes up about five sixths of the matter in the Universe, yet we still do not know what it is~\cite{bertone2005}. NSs can capture DM from the Galactic halo~\cite{kouvaris2010dark}, but capture alone gives DM fractions far below the percent level~\cite{Kalita2026}. Larger fractions are possible if DM is already mixed in when the progenitor forms in a DM-rich region, if it is produced inside the star, or if it is taken from a dark compact companion~\cite{Ivanytskyi2020,karkevandi2022}. In each case the DM forms a second fluid that interacts with ordinary matter mainly through gravity. Hyperons and quarks change the composition of the baryonic sector, whereas the case for a DM component has nothing to do with nuclear physics. Its effect on the star can nevertheless look very much like a change in the nuclear EoS, and telling the two apart is part of reading NS data correctly.

Sandin and Ciarcelluti~\cite{sandin2009dark} showed that a small admixture of mirror matter is enough to reshape the $M$--$R$ relation, and Ciarcelluti and Sandin~\cite{ciarcelluti2011} asked whether NSs could carry a DM core. Leung, Chu and Lin~\cite{leung2011} set up the general-relativistic two-fluid problem for DM-admixed neutron stars (DANSs) and found two kinds of solutions: a compact DM core inside the baryonic star, and a small baryonic star embedded in an extended DM halo. Later work examined the radial stability and oscillation modes of these objects~\cite{leung2012} and the role of the DM particle mass and self-interaction~\cite{xiang2014}. The same two-fluid picture was then applied to tidal effects. Das, Malik and Nayak~\cite{das2019,das2022dark} studied how DM changes the structure and tidal deformability, and Leung, Chu and Lin~\cite{leung2022} computed the Love number for both fermionic and bosonic DM. A rough picture has emerged from these studies: halos tend to increase $\Lambda$ and the gravitational mass~\cite{Nelson2019,karkevandi2022,Giangrandi2023}, while cores reduce both~\cite{Ivanytskyi2020,das2022dark,Routaray2023}. More recently, Rutherford et~al.~\cite{Rutherford2025} carried out Bayesian analyses of NICER data and found that stars with fermionic asymmetric-DM cores fit the current mass--radius measurements as well as purely baryonic stars do. For DM fractions up to $1.7\%$ the inferences with and without DM are almost the same, so mass--radius data by themselves say little about a DM core~\cite{Rutherford2025,Rutherford2026}.

This is the starting point of the present work. We ask how a fermionic DM core, coupled to nuclear matter only through gravity, changes not only the $M$--$R$ relation but also the compactness, Love number, tidal deformability and moment of inertia. For nucleonic matter we use the HPU-3 relativistic mean-field (RMF) EoS~\cite{kumar2023observational}, which can support the mass of PSR~J0952$-$0607 and so leaves room for a DM core below the observed maximum masses. For DM we use a Walecka-type model with scalar and vector self-interactions~\cite{das2022dark}. We consider DM mass fractions up to $\FDM=10\%$, follow the radial distribution of both fluids, and compare the results with GW170817, GW190425, NICER and radio-timing constraints.

Our additions to earlier two-fluid studies are the following. First, we work out how much DM a stiff EoS of this kind can hold, and find that a core of $3$--$4\%$ is enough to bring it within the GW170817 tidal bounds. We present this as a degeneracy between the DM content and the nuclear EoS, not as evidence for DM, because adding DM can make a nuclear EoS that fails the tidal test on its own look acceptable~\cite{Kalita2026}. Second, we obtain the moment of inertia from the slow-rotation equations and show that the widely used empirical estimate is not accurate enough to resolve the effect of DM. Third, we explain the formation of the core with a simple pseudo-enthalpy argument and test how the results depend on the DM particle mass and couplings and on the stiffness of the nuclear EoS, by repeating the calculation with the softer SLy EoS. Rotation, radial stability and oscillation modes of two-fluid stars have been studied recently in Refs.~\cite{Cronin2023,Caballero2024,KumarSotani2025,KumarSotani2026,SotaniKumar2025}, and the boundary between cores and halos across DM masses in Ref.~\cite{Karan2026}. We compare with these works where they overlap with ours.

The paper is organized as follows. Section~\ref{sec:theory} introduces the two EoSs, their analytic representation, and the equations for the stellar structure and its perturbations. Section~\ref{sec:results} presents the results, and Sec.~\ref{sec:conclusions} sums up. The model parameters are collected in Appendix~\ref{app:params}, the numerical method and its tests in Appendix~\ref{app:verification}, and supplementary figures in Appendix~\ref{app:figures}. Equations are written in units with $G=c=1$ unless noted otherwise, while numerical results are given in physical units.

\section{Theoretical framework}\label{sec:theory}

We write $\varepsilon$ for energy density, $P$ for pressure and $n$ for number density, and keep the symbol $\rho$ for the isovector meson field. In figures and tables, energy densities are quoted as $\varepsilon/c^2$ in $\gcm$. Sections~\ref{sec:hpu3} and \ref{sec:fdm} use the particle-physics metric signature $\eta_{\mu\nu}={\rm diag}(1,-1,-1,-1)$ with $\hbar=c=1$. The stellar-structure equations in Secs.~\ref{sec:structure}--\ref{sec:tidal} use the signature $(-,+,+,+)$ with $G=c=1$.

\subsection{Nuclear matter: the HPU-3 equation of state}\label{sec:hpu3}

For $\beta$-equilibrated nucleonic matter we use the HPU-3 parametrization of the RMF model. Kumar et~al.~\cite{kumar2023observational} built it to support the mass of PSR~J0952$-$0607 while still meeting finite-nuclei and bulk nuclear-matter constraints. In this model nucleons behave as relativistic quasiparticles moving in classical meson mean fields~\cite{Walecka1974,SerotWalecka1986}. The isoscalar-scalar $\sigma$ meson gives the intermediate-range attraction and the isoscalar-vector $\omega$ meson the short-range repulsion. The isovector-vector $\rho$ meson couples to the isospin asymmetry and so controls how the symmetry energy changes with density. The Lagrangian density is
\begin{widetext}
\begin{equation}
\begin{split}
\mathcal{L}=&\sum_{B}\bar{\psi}_{B}\Big[i\gamma^{\mu}\partial_{\mu}
-\left(M_{B}-g_{\sigma}\sigma\right)
-g_{\omega}\gamma^{\mu}\omega_{\mu}
-\frac{1}{2}g_{\rho}\gamma^{\mu}\bm{\tau}_{B}\cdot\bm{\rho}_{\mu}\Big]\psi_{B}
+\frac{1}{2}\left(\partial_{\mu}\sigma\,\partial^{\mu}\sigma-m_{\sigma}^{2}\sigma^{2}\right)
\\
&-\frac{\kappa}{3!}\left(g_{\sigma}\sigma\right)^{3}
-\frac{\lambda}{4!}\left(g_{\sigma}\sigma\right)^{4}
-\frac{1}{4}\omega_{\mu\nu}\omega^{\mu\nu}
+\frac{1}{2}m_{\omega}^{2}\omega_{\mu}\omega^{\mu}
+\frac{\zeta}{4!}g_{\omega}^{4}\left(\omega_{\mu}\omega^{\mu}\right)^{2}
-\frac{1}{4}\bm{\rho}_{\mu\nu}\cdot\bm{\rho}^{\mu\nu}
+\frac{1}{2}m_{\rho}^{2}\bm{\rho}_{\mu}\cdot\bm{\rho}^{\mu}
+\frac{\xi}{4!}g_{\rho}^{4}\left(\bm{\rho}_{\mu}\cdot\bm{\rho}^{\mu}\right)^{2}
\\
&+g_{\sigma}g_{\omega}^{2}\sigma\,\omega_{\mu}\omega^{\mu}\left(a_{1}+\frac{1}{2}a_{2}g_{\sigma}\sigma\right)
+g_{\sigma}g_{\rho}^{2}\sigma\,\bm{\rho}_{\mu}\cdot\bm{\rho}^{\mu}\left(b_{1}+\frac{1}{2}b_{2}g_{\sigma}\sigma\right)
\\
&+\frac{1}{2}\Lambda_{v}g_{\omega}^{2}g_{\rho}^{2}\,\omega_{\mu}\omega^{\mu}\,\bm{\rho}_{\nu}\cdot\bm{\rho}^{\nu}
+\sum_{\ell=e,\mu}\bar{\psi}_{\ell}\left(i\gamma^{\mu}\partial_{\mu}-m_{\ell}\right)\psi_{\ell},
\end{split}
\label{eq:RMF_Lagrangian}
\end{equation}
\end{widetext}
where $B\in\{n,p\}$, $\omega_{\mu\nu}=\partial_\mu\omega_\nu-\partial_\nu\omega_\mu$, and $\bm{\rho}_{\mu\nu}$ is defined in the same way. We leave out the electromagnetic field, which plays no role in uniform, charge-neutral matter. The field equations follow from the Euler--Lagrange equations, and the energy--momentum tensor is
\begin{equation}
T_{\mu\nu}=\sum_{\varphi}\frac{\partial\mathcal{L}}{\partial(\partial^{\mu}\varphi)}\partial_{\nu}\varphi-\eta_{\mu\nu}\mathcal{L}.
\label{eq:Tmunu}
\end{equation}
In uniform matter at rest the meson fields can be replaced by their mean values $\sigma$, $\omega\equiv\langle\omega_0\rangle$ and $\rho\equiv\langle\rho_{0}^{3}\rangle$, and the energy density and pressure are $\varepsilon=\langle T_{00}\rangle$ and $P=\frac13\sum_i\langle T_{ii}\rangle$~\cite{Glendenning2000}. With the Fermi momentum $k_B$, the number density $n_B=k_B^3/(3\pi^2)$ and the effective mass $M_B^{*}=M_B-g_\sigma\sigma$, the energy density and pressure of nuclear matter (NM) take the form
\begin{widetext}
\begin{equation}
\begin{split}
\varepsilon_{\NM}=&
\sum_{B}\frac{1}{\pi^{2}}\int_{0}^{k_{B}}k^{2}\sqrt{k^{2}+M_{B}^{*2}}\,dk
+\sum_{\ell}\frac{1}{\pi^{2}}\int_{0}^{k_{\ell}}k^{2}\sqrt{k^{2}+m_{\ell}^{2}}\,dk
\\
&+g_{\omega}\omega\, n
+\frac{1}{2}g_{\rho}\rho\sum_{B}\tau_{3B}n_{B}
+\frac{1}{2}m_{\sigma}^{2}\sigma^{2}
+\frac{\kappa}{6}(g_{\sigma}\sigma)^{3}
+\frac{\lambda}{24}(g_{\sigma}\sigma)^{4}
\\
&-\frac{1}{2}m_{\omega}^{2}\omega^{2}
-\frac{\zeta}{24}g_{\omega}^{4}\omega^{4}
-\frac{1}{2}m_{\rho}^{2}\rho^{2}
-\frac{\xi}{24}g_{\rho}^{4}\rho^{4}
\\
&-g_{\sigma}g_{\omega}^{2}\sigma\omega^{2}\Big(a_{1}+\frac{1}{2}a_{2}g_{\sigma}\sigma\Big)
-g_{\sigma}g_{\rho}^{2}\sigma\rho^{2}\Big(b_{1}+\frac{1}{2}b_{2}g_{\sigma}\sigma\Big)
\\
&-\frac{1}{2}\Lambda_{v}g_{\omega}^{2}g_{\rho}^{2}\omega^{2}\rho^{2},
\end{split}
\label{eq:eps_NM}
\end{equation}
\begin{equation}
\begin{split}
P_{\NM}=&
\sum_{B}\frac{1}{3\pi^{2}}\int_{0}^{k_{B}}\frac{k^{4}\,dk}{\sqrt{k^{2}+M_{B}^{*2}}}
+\sum_{\ell}\frac{1}{3\pi^{2}}\int_{0}^{k_{\ell}}\frac{k^{4}\,dk}{\sqrt{k^{2}+m_{\ell}^{2}}}
\\
&-\frac{1}{2}m_{\sigma}^{2}\sigma^{2}
-\frac{\kappa}{6}(g_{\sigma}\sigma)^{3}
-\frac{\lambda}{24}(g_{\sigma}\sigma)^{4}
+\frac{1}{2}m_{\omega}^{2}\omega^{2}
+\frac{\zeta}{24}g_{\omega}^{4}\omega^{4}
\\
&+\frac{1}{2}m_{\rho}^{2}\rho^{2}
+\frac{\xi}{24}g_{\rho}^{4}\rho^{4}
+g_{\sigma}g_{\omega}^{2}\sigma\omega^{2}\Big(a_{1}+\frac{1}{2}a_{2}g_{\sigma}\sigma\Big)
\\
&+g_{\sigma}g_{\rho}^{2}\sigma\rho^{2}\Big(b_{1}+\frac{1}{2}b_{2}g_{\sigma}\sigma\Big)
+\frac{1}{2}\Lambda_{v}g_{\omega}^{2}g_{\rho}^{2}\omega^{2}\rho^{2},
\end{split}
\label{eq:P_NM}
\end{equation}
\end{widetext}
\noindent where $n=n_n+n_p$ and $\tau_{3B}=\pm1$ for protons and neutrons. The mean fields follow from extremizing $\varepsilon_{\NM}$ at fixed densities. The composition is set by charge neutrality, $n_p=n_e+n_\mu$, and by $\beta$ equilibrium, $\mu_n=\mu_p+\mu_e$ and $\mu_\mu=\mu_e$. Because the meson fields obey their equations of motion, the chemical potentials are simply $\mu_i=\partial\varepsilon_{\NM}/\partial n_i$, and Eq.~\eqref{eq:P_NM} agrees with the thermodynamic pressure
\begin{equation}
P_{\NM}=\sum_{i}\mu_{i}n_{i}-\varepsilon_{\NM},
\label{eq:euler_relation}
\end{equation}
where $i$ runs over baryons and leptons. The model is therefore thermodynamically consistent~\cite{Glendenning2000}, and for symmetric matter at saturation Eq.~\eqref{eq:euler_relation} reduces to the Hugenholtz--Van Hove theorem~\cite{Hugenholtz1958}. The HPU-3 couplings are listed in Table~\ref{tab:HPU3parameters} of Appendix~\ref{app:params}. At saturation the model gives $n_0=0.151\,\mathrm{fm}^{-3}$, $E/A=-16.06$~MeV, $K=225.6$~MeV, $M^{*}/M=0.635$, $J=33.2$~MeV, $L=75.0$~MeV and $K_{\rm sym}=-31.4$~MeV, and a neutron-skin thickness of $0.217$~fm for $^{208}$Pb~\cite{kumar2023observational}. The slope $L$ lies inside the broad range of current estimates but above some recent determinations, which fits the rather large radius $R_{1.4}\approx13.4$~km found below. We do not impose chiral effective-field-theory constraints at low density. As in Ref.~\cite{kumar2023observational}, the crust is described by the BPS EoS~\cite{BPS1971} joined to the RMF EoS at $0.5\,n_0$, so one analytic representation (Sec.~\ref{sec:fit}) describes the star all the way from the surface to the centre.

\subsection{Fermionic dark matter}\label{sec:fdm}

We follow Das et~al.~\cite{das2022dark} and take the DM to be a single Dirac fermion $\psi_D$ of mass $M_D$. Its self-interaction is carried by a dark scalar $\phi$, which attracts, and a dark vector $V_\mu$, which repels, much like the $\sigma$ and $\omega$ mesons in nuclear matter. Keeping to one fermion avoids adding several dark species whose properties are poorly constrained. Bosonic and mirror DM, which can behave quite differently inside neutron stars, are outside the scope of this work. The Lagrangian density is
\begin{equation}
\begin{split}
\mathcal{L}_{\DM}=&
\bar{\psi}_D\gamma^{\mu}\left(i\partial_{\mu}-g_{vD}V_{\mu}\right)\psi_D
-\bar{\psi}_D\left(M_{D}-g_{sD}\phi\right)\psi_D
\\
&+\frac{1}{2}\left(\partial^{\mu}\phi\,\partial_{\mu}\phi-m_{sD}^{2}\phi^{2}\right)
\\
&-\frac{1}{4}V_{\mu\nu}V^{\mu\nu}
+\frac{1}{2}m_{vD}^{2}V^{\mu}V_{\mu}.
\end{split}
\label{eq:DM_Lagrangian}
\end{equation}
In the mean-field approximation the DM energy density is the sum of a relativistic Fermi-gas term and the scalar and vector mean-field terms,
\begin{equation}
\begin{split}
\varepsilon_{\DM}=&
\frac{1}{\pi^{2}}\int_{0}^{k_{D}}k^{2}\sqrt{k^{2}+M_{D}^{*2}}\,dk
+\frac{1}{2}\left(\frac{g_{vD}}{m_{vD}}\right)^{2}n_{D}^{2}
\\
&+\frac{1}{2}\left(\frac{m_{sD}}{g_{sD}}\right)^{2}\left(M_{D}-M_{D}^{*}\right)^{2},
\end{split}
\label{eq:DM_energy_density}
\end{equation}
and the pressure is
\begin{equation}
\begin{split}
P_{\DM}=&
\frac{1}{3\pi^{2}}\int_{0}^{k_{D}}\frac{k^{4}\,dk}{\sqrt{k^{2}+M_{D}^{*2}}}
+\frac{1}{2}\left(\frac{g_{vD}}{m_{vD}}\right)^{2}n_{D}^{2}
\\
&-\frac{1}{2}\left(\frac{m_{sD}}{g_{sD}}\right)^{2}\left(M_{D}-M_{D}^{*}\right)^{2}.
\end{split}
\label{eq:DM_pressure}
\end{equation}
The vector field pushes the pressure up and the scalar field pulls it down, and the balance between the two sets how stiff the DM EoS is. The DM number density and effective mass are
\begin{align}
n_D&=\frac{k_D^{3}}{3\pi^{2}},
\label{eq:DM_number_density}\\
M_D^{*}&=M_D-\left(\frac{g_{sD}}{m_{sD}}\right)^{2}n_{sD},
\label{eq:DM_effective_mass}\\
n_{sD}&=\frac{1}{\pi^{2}}\int_{0}^{k_D}\frac{k^{2}M_D^{*}}{\sqrt{k^{2}+M_D^{*2}}}\,dk,
\label{eq:DM_scalar_density}
\end{align}
where Eq.~\eqref{eq:DM_effective_mass} is solved self-consistently for each $k_D$. The EoS depends only on $M_D$ and on the ratios $g_{sD}/m_{sD}$ and $g_{vD}/m_{vD}$, whose values are listed in Table~\ref{tab:dm_couplings}. They are the posterior medians of the Bayesian fit of Ref.~\cite{das2022dark}, which was made with a different nuclear EoS, so we use them as a representative benchmark and test the sensitivity to them in Sec.~\ref{sec:sensitivity}. In the limit of heavy mediators the same ratios also set the DM self-scattering cross-section. With the Born expression used in Ref.~\cite{DasWei2026}, the vector coupling gives $\sigma/M_D\approx0.25\,\mathrm{cm^2\,g^{-1}}$, or about $0.20\,\mathrm{cm^2\,g^{-1}}$ once the scalar attraction is subtracted. Both values lie below the Bullet Cluster limit $\sigma/m<1.25\,\mathrm{cm^2\,g^{-1}}$~\cite{Randall2008}.

\subsection{Analytic representation of the equations of state}\label{sec:fit}

To solve the structure equations we need $P(\varepsilon)$ and its derivative everywhere between the surface and the centre, without jumps. Instead of interpolating tables, we represent each EoS by a smooth analytic function, following the idea of Refs.~\cite{HaenselPotekhin2004,Potekhin2013,Read2009}. Pressure and energy density span many orders of magnitude, and nuclear EoSs behave locally like power laws, so the fit is made in logarithmic variables,
\begin{equation}
\xi=\log_{10}\!\left(\frac{\varepsilon/c^{2}}{\gcm}\right),\qquad
\zeta=\log_{10}\!\left(\frac{P}{\mathrm{dyn\,cm^{-2}}}\right).
\label{eq:xi_zeta}
\end{equation}
With the switching function $f_0(x)=1/(e^{x}+1)$ we write
\begin{align}
\zeta_{\rm low}(\xi)&=\left[c_{1}+c_{2}\left|\xi-c_{3}\right|^{c_{4}}\right]f_{0}\!\left(c_{5}(\xi-c_{6})\right)
\nonumber\\
&\quad+\left(c_{7}+c_{8}\xi\right)f_{0}\!\left(c_{9}(c_{10}-\xi)\right),
\label{eq:zeta_low}\\
\zeta_{\rm high}(\xi)&=\left(\alpha_{3}+\alpha_{4}\xi\right)f_{0}\!\left(\alpha_{5}(\alpha_{6}-\xi)\right)
\nonumber\\
&\quad+\left(\alpha_{7}+\alpha_{8}\xi+\alpha_{9}\xi^{2}\right)f_{0}\!\left(\alpha_{10}(\alpha_{11}-\xi)\right),
\label{eq:zeta_high}\\
\zeta(\xi)&=\zeta_{\rm low}(\xi)\,f_{0}\!\left(\alpha_{1}(\xi-c_{11})\right)
\nonumber\\
&\quad+\zeta_{\rm high}(\xi)\,f_{0}\!\left(\alpha_{2}(c_{12}-\xi)\right).
\label{eq:zeta_total}
\end{align}
The coefficients are listed in Table~\ref{tab:fitcoeffs}. The HPU-3 fit covers $\xi\geq4.02$, crust included; below that density we set $P=0$, which defines the stellar surface. The DM fit covers $9.11\leq\xi\leq17.07$. At lower densities the Walecka EoS approaches a non-relativistic Fermi gas, so there we continue it as $P\propto\varepsilon^{5/3}$, joined continuously in $P$. When we evaluate Eqs.~\eqref{eq:DM_energy_density}--\eqref{eq:DM_scalar_density} directly with the parameters of Table~\ref{tab:dm_couplings}, we recover the fitted $\zeta$ to within $10^{-3}$ (0.2\% in $P$) across the fitted range (Appendix~\ref{app:verification}). Both representations are monotonic and causal ($0<c_s^2<1$) at all densities reached in our stars (Fig.~\ref{fig:eos}).

\subsection{Two-fluid stellar structure}\label{sec:structure}

Inside a static, spherically symmetric star we use the metric
\begin{equation}
\begin{split}
ds^{2}=&-e^{2\nu(r)}dt^{2}+\left(1-\frac{2m(r)}{r}\right)^{-1}dr^{2}
\\
&+r^{2}\left(d\theta^{2}+\sin^{2}\theta\,d\phi^{2}\right),
\end{split}
\label{eq:metric}
\end{equation}
where $m(r)$ is the mass--energy inside radius $r$. Outside the star the metric is the Schwarzschild solution, so $e^{2\nu(R)}=1-2M/R$. For a single perfect fluid, the Einstein equations together with $\nabla_\mu T^{\mu\nu}=0$ give the Tolman--Oppenheimer--Volkoff (TOV) equations~\cite{tolman1939,oppenheimer1939},
\begin{align}
\frac{dm}{dr}&=4\pi r^{2}\varepsilon,\label{eq:tov-m}\\
\frac{dP}{dr}&=-(\varepsilon+P)\frac{d\nu}{dr},\qquad
\frac{d\nu}{dr}=\frac{m+4\pi r^{3}P}{r\left(r-2m\right)}.\label{eq:tov-p}
\end{align}
We start at the centre with $m(0)=0$ and $P(0)=P_c$ and integrate outward until $P(R)=0$. This defines the radius $R$ and the gravitational mass $M=m(R)$.

If the star is made of nuclear matter and DM that feel each other only through gravity, each fluid obeys its own conservation law, $\nabla_\mu T_i^{\mu\nu}=0$, while both act as sources of the same metric~\eqref{eq:metric}~\cite{sandin2009dark,ciarcelluti2011,leung2011}. With $\varepsilon=\varepsilon_{\NM}+\varepsilon_{\DM}$ and $P=P_{\NM}+P_{\DM}$, the two-fluid TOV equations are
\begin{align}
\frac{dm_i}{dr}&=4\pi r^{2}\varepsilon_i,\qquad m=m_{\NM}+m_{\DM},\label{eq:tov2-m}\\
\frac{dP_i}{dr}&=-(\varepsilon_i+P_i)\,\frac{m+4\pi r^{3}(P_{\NM}+P_{\DM})}{r\left(r-2m\right)},\label{eq:tov2-p}
\end{align}
for $i\in\{\NM,\DM\}$. Each fluid has its own EoS $P_i(\varepsilon_i)$ and its own surface, where $P_i(R_i)=0$. The total gravitational mass is $M=M_{\NM}+M_{\DM}$, with $M_i=m_i(R_{\rm out})$ and $R_{\rm out}=\max(R_{\NM},R_{\DM})$. The DM mass fraction is
\begin{equation}
\FDM=\frac{M_{\DM}}{M_{\NM}+M_{\DM}}.
\label{eq:FDM}
\end{equation}

We fix the mass fraction rather than the number of DM particles. The two differ through the internal and gravitational binding energy of each fluid, which changes the numbers but not the trends. The fraction $f_D$ of Ref.~\cite{das2022dark} is instead the ratio of the central energy densities, so their posterior value $f_D=0.08$ does not correspond to $\FDM=8\%$.

Where the DM sits matters as much as how much of it there is. We speak of a DM core when $R_{\DM}\leq R_{\NM}$ and of a DM halo when $R_{\DM}>R_{\NM}$, with both radii taken for the same star~\cite{leung2011,karkevandi2022}. For a core, the visible radius measured in X-rays is also the outermost radius, $R=R_{\NM}=R_{\rm out}$.

We label a stellar sequence by the total central energy density $\varepsilon_c=\varepsilon_{\NM}(0)+\varepsilon_{\DM}(0)$. For each $\varepsilon_c$ and each target value $\FDM\in\{0,0.1,1,5,10\}\%$, a root search adjusts the ratio $\varepsilon_{\DM}(0)/\varepsilon_c$ until Eq.~\eqref{eq:FDM} holds to $10^{-9}$. We use 70 values between $\varepsilon_c/c^2=10^{14.3}$ and $10^{15.8}\,\gcm$. The maximum of $M(\varepsilon_c)$ along each fixed-$\FDM$ sequence is located with a bounded one-dimensional search, and we take it as the onset of instability~\cite{Harrison:1965gtgc}. For two fluids this is only an approximation, since a strict criterion has to consider the two-parameter family of central densities at fixed particle numbers~\cite{Henriques:1990xg,leung2012}. The radial-mode basis of this criterion has been established in Ref.~\cite{Caballero2024}, and for two-fluid DM-admixed stars the boundaries found from radial perturbations and from a generalized turning-point criterion agree closely~\cite{KumarSotani2025}. We therefore do not repeat a radial-mode analysis here.

We integrate Eqs.~\eqref{eq:tov2-m}--\eqref{eq:tov2-p} in pseudo-enthalpy form~\cite{Lindblom1992}. For each fluid, $h_i=\int_0^{P_i}dP/(\varepsilon_i+P)$ satisfies $dh_i/dr=-d\nu/dr$. Both fluids can therefore use the same independent variable $t=h_i(0)-h_i(r)$, and fluid $i$ ends exactly at $t=h_i(0)$. In this way we need neither density cutoffs nor numerical derivatives $dP/d\varepsilon$, and both surfaces are located exactly. Details and tests are given in Appendix~\ref{app:verification}.

\subsection{Tidal deformability}\label{sec:tidal}

In a binary, the tidal field $\mathcal{E}_{ij}$ of the companion induces a mass quadrupole moment $Q_{ij}$ in the star. At leading order~\cite{flanagan2008constraining,hinderer2008tidal}
\begin{equation}
Q_{ij}=-\lambda\,\mathcal{E}_{ij},\qquad
\lambda=\frac{2}{3}k_2R^{5},\qquad
\Lambda\equiv\frac{\lambda}{M^{5}}=\frac{2}{3}k_2\,C^{-5},
\label{eq:Lambda}
\end{equation}
where $k_2$ is the quadrupolar Love number (the idea goes back to Love's work on the Earth~\cite{Love:1911spge}), $C=M/R$ is the compactness, and $R=R_{\rm out}$. Tidal deformation speeds up the inspiral and leaves a measurable phase shift in the gravitational-wave signal~\cite{flanagan2008constraining}, which LIGO and Virgo used to constrain the EoS with GW170817~\cite{abbott2017gw170817,Abbott2018}.

The Love number comes from static, even-parity ($\ell=2$) perturbations of the metric~\eqref{eq:metric}~\cite{hinderer2008tidal,DamourNagar2009,Hinderer2010,postnikov2010tidal}. Writing the metric perturbation as $H(r)Y_{2m}$ and defining $y=rH'/H$ turns the second-order equation for $H$ into a first-order Riccati equation,
\begin{equation}
r\frac{dy}{dr}+y^{2}+y\,e^{\lambda}\Big[1+4\pi r^{2}(P-\varepsilon)\Big]+r^{2}Q(r)=0,
\label{eq:y-ode}
\end{equation}
with $e^{\lambda}=(1-2m/r)^{-1}$ and
\begin{equation}
Q(r)=4\pi e^{\lambda}\Bigg[5\varepsilon+9P+\sum_{i}\frac{\varepsilon_i+P_i}{dP_i/d\varepsilon_i}\Bigg]-\frac{6e^{\lambda}}{r^{2}}-4\left(\frac{d\nu}{dr}\right)^{2}.
\label{eq:Q}
\end{equation}
The only difference from the single-fluid case is the sum over fluids in the term that contains the sound speeds $dP_i/d\varepsilon_i$~\cite{das2022dark,leung2022}. The factor 4 in front of $(d\nu/dr)^2$ appears because our $\nu$ in Eq.~\eqref{eq:metric} is half of the metric function used in Refs.~\cite{hinderer2008tidal,postnikov2010tidal}. Equation~\eqref{eq:y-ode} is regular at the centre, where $y(0)=2$, and we integrate it out to the outer surface to obtain $y_R\equiv y(R)$. Matching to the exterior solution then gives~\cite{hinderer2008tidal,DamourNagar2009,postnikov2010tidal}
\begin{equation}
k_2=\frac{8C^5}{5}\,\frac{(1-2C)^2\left[2+2C(y_R-1)-y_R\right]}{x_1+x_2+x_3},
\label{eq:k2}
\end{equation}
with
\begin{align}
x_1&=2C\left[6-3y_R+3C(5y_R-8)\right],\nonumber\\
x_2&=4C^3\left[13-11y_R+C(3y_R-2)+2C^2(1+y_R)\right],\label{eq:k2aux}\\
x_3&=3(1-2C)^2\left[2-y_R+2C(y_R-1)\right]\ln(1-2C).\nonumber
\end{align}
The compactness tells us how relativistic the star is, while $y_R$ carries the information about its internal density profile. Both fluids reach zero density at their surfaces, and HPU-3 ends at $\varepsilon/c^2\approx10^{4}\,\gcm$, so the surface-density correction to $y_R$~\cite{DamourNagar2009,postnikov2010tidal} can be neglected. For a binary with masses $m_1\geq m_2$, the waveform depends at leading order on~\cite{flanagan2008constraining,Abbott2019PRX}
\begin{equation}
\tilde\Lambda=\frac{16}{13}\frac{(m_1+12m_2)\,m_1^{4}\Lambda_1+(m_2+12m_1)\,m_2^{4}\Lambda_2}{(m_1+m_2)^{5}}.
\label{eq:Lambdatilde}
\end{equation}
For GW170817 we take the source-frame chirp mass $1.186\,\Msun$ and mass ratios $0.73\leq m_2/m_1\leq1$ (low-spin prior)~\cite{Abbott2019PRX}, and assume that both stars have the same $\FDM$.

\section{Results and discussion}\label{sec:results}

Figure~\ref{fig:eos} in Appendix~\ref{app:figures} compares the two EoSs. At the same energy density the DM EoS is stiffer than HPU-3 between about $8\times10^{11}$ and $1.5\times10^{15}\,\gcm$, so relative stiffness alone does not decide whether the DM forms a core or a halo. The pseudo-enthalpy form of Sec.~\ref{sec:structure} gives a cleaner answer. Both enthalpies fall outward at the same rate, $dh_i/dr=-d\nu/dr$, so the fluid that starts with the smaller central enthalpy runs out first. In our stars the DM is the minority component, its central energy density is several times smaller than that of nuclear matter, and $h_{\DM}(0)<h_{\NM}(0)$. In a $1.4\,\Msun$ star, for example, $h_{\DM}(0)=0.034$ and $h_{\NM}(0)=0.195$ at $\FDM=0.1\%$, and $0.197$ and $0.223$ at $10\%$. The DM therefore ends inside the nuclear matter. A lighter DM particle has a larger enthalpy at the same energy density, so for light DM the DM can extend beyond the nuclear matter and form a halo, as we show in Sec.~\ref{sec:sensitivity}. Table~\ref{tab:mmax} lists the maximum-mass configurations and Table~\ref{tab:canonical} the properties of $1.4\,\Msun$ stars.

\begin{table*}[t]
\caption{Maximum-mass configurations for each DM fraction: gravitational mass $M_{\max}$, nuclear-matter radius $R$, DM-core radius $R_{\DM}$, total central energy density $\varepsilon_c$, fraction of the central energy density carried by DM, compactness $C$, Love number $k_2$, tidal deformability $\Lambda$, moment of inertia $I$ from the slow-rotation equations, and the Ravenhall--Pethick estimate $I_{\rm RP}$ of Eq.~\eqref{eq:I_RP}.}
\label{tab:mmax}
\begin{ruledtabular}
\begin{tabular}{lcccccccccc}
$\FDM$ & $M_{\max}\,[\Msun]$ & $R\,[\mathrm{km}]$ & $R_{\DM}\,[\mathrm{km}]$ & $\varepsilon_c/c^2\,[10^{15}\gcm]$ & $\varepsilon_{\DM}(0)/\varepsilon_c$ & $C$ & $k_2$ & $\Lambda$ & $I\,[10^{45}\mathrm{g\,cm^2}]$ & $I_{\rm RP}\,[10^{45}\mathrm{g\,cm^2}]$ \\
\colrule
$0$      & 2.351 & 12.00 & ---  & 1.945 & ---   & 0.2893 & 0.0225 & 7.40 & 3.079 & 3.353 \\
$0.1\%$  & 2.347 & 11.99 & 3.82 & 2.014 & 0.030 & 0.2890 & 0.0225 & 7.43 & 3.068 & 3.341 \\
$1\%$    & 2.320 & 11.93 & 5.81 & 2.171 & 0.084 & 0.2873 & 0.0225 & 7.65 & 2.974 & 3.240 \\
$5\%$    & 2.218 & 11.65 & 7.71 & 2.470 & 0.173 & 0.2812 & 0.0224 & 8.47 & 2.630 & 2.871 \\
$10\%$   & 2.115 & 11.32 & 8.71 & 2.662 & 0.235 & 0.2760 & 0.0222 & 9.23 & 2.306 & 2.525 \\
\end{tabular}
\end{ruledtabular}
\end{table*}

\begin{table*}[t]
\caption{Properties of $1.4\,\Msun$ stars and related quantities: radius $R_{1.4}$, DM-core radius, compactness, Love number, tidal deformability, moment of inertia, radius of a $2.08\,\Msun$ star (the mass of PSR~J0740+6620), and the range of the binary tidal deformability $\tilde\Lambda$, Eq.~\eqref{eq:Lambdatilde}, for the chirp mass of GW170817 ($1.186\,\Msun$) and mass ratios $0.73\leq q\leq1$~\cite{Abbott2019PRX}.}
\label{tab:canonical}
\begin{ruledtabular}
\begin{tabular}{lcccccccc}
$\FDM$ & $R_{1.4}\,[\mathrm{km}]$ & $R_{\DM,1.4}\,[\mathrm{km}]$ & $C_{1.4}$ & $k_{2,1.4}$ & $\Lambda_{1.4}$ & $I_{1.4}\,[10^{45}\mathrm{g\,cm^2}]$ & $R_{2.08}\,[\mathrm{km}]$ & $\tilde\Lambda$ (GW170817) \\
\colrule
$0$      & 13.44 & ---   & 0.1538 & 0.0939 & 728 & 1.729 & 13.11 & 837--853 \\
$0.1\%$  & 13.43 & 5.00  & 0.1539 & 0.0936 & 723 & 1.726 & 13.09 & 831--848 \\
$1\%$    & 13.35 & 7.57  & 0.1549 & 0.0913 & 684 & 1.700 & 12.98 & 787--802 \\
$5\%$    & 13.02 & 10.08 & 0.1588 & 0.0847 & 558 & 1.608 & 12.49 & 649--658 \\
$10\%$   & 12.64 & 11.46 & 0.1635 & 0.0808 & 461 & 1.529 & 11.78 & 545--547 \\
\end{tabular}
\end{ruledtabular}
\end{table*}

\subsection{Mass--radius and mass--central-density relations}\label{sec:MR}

\begin{figure*}[t]
\centering
\includegraphics[width=\textwidth]{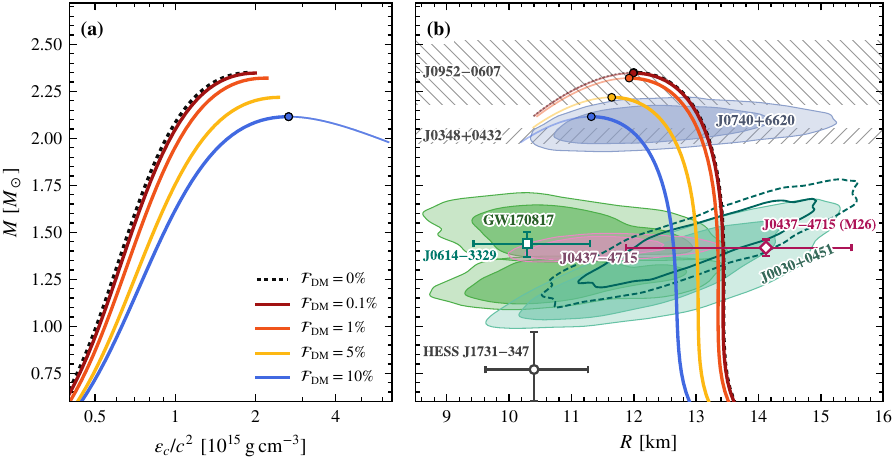}
\caption{(a) Gravitational mass versus total central energy density $\varepsilon_c$ for $\FDM=0$, $0.1$, $1$, $5$ and $10\%$. Thick lines are the stable branches, thin faded lines the configurations beyond the maximum mass, and circles the maximum-mass stars. (b) Mass--radius relations for the same sequences, on the same mass axis; $R$ is the radius of the nuclear-matter surface, and only configurations with a DM core are shown. Shaded regions are the 68\% (darker) and 90\% (lighter) credible regions of GW170817 (both components, EoS-insensitive analysis~\cite{Abbott2018}), PSR~J0030+0451 (filled: Ref.~\cite{riley2019}; solid and dashed outlines: 68\% and 90\% of Ref.~\cite{miller2019nicer}), PSR~J0437$-$4715~\cite{Choudhury:2024xbk} and PSR~J0740+6620~\cite{Riley2021}. Points with error bars show 68\% intervals for PSR~J0614$-$3329~\cite{Mauviard2025}, the reanalysis of PSR~J0437$-$4715 labelled M26~\cite{Miller2026a}, and HESS~J1731$-$347~\cite{Doroshenko2022}. Hatched bands mark the $1\sigma$ masses of PSR~J0348+0432~\cite{antoniadis2013} and PSR~J0952$-$0607~\cite{romani2022}.}
\label{fig:MR}
\end{figure*}

Figure~\ref{fig:MR}(a) shows how the mass depends on the total central energy density, and Fig.~\ref{fig:MR}(b) shows the $M$--$R$ relations; the two panels share the mass axis. Without DM, HPU-3 reaches $M_{\max}=2.351\,\Msun$ at $R=12.00$~km, and a $1.4\,\Msun$ star has $R_{1.4}=13.44$~km. These numbers agree with the HPU-3 values of Ref.~\cite{kumar2023observational} to within $0.4\%$ in radius (Appendix~\ref{app:verification}), so the analytic fit of Sec.~\ref{sec:fit} keeps the macroscopic properties of the original EoS.

A small amount of DM hardly changes anything. At $\FDM=0.1\%$ the maximum mass goes down by $0.003\,\Msun$ and $R_{1.4}$ by $0.01$~km, and even at $\FDM=1\%$ the shifts stay modest ($M_{\max}=2.320\,\Msun$, $R_{1.4}=13.35$~km). Larger fractions have a clear effect. For $\FDM=5\%$ and $10\%$ the maximum mass falls to $2.218$ and $2.115\,\Msun$, the corresponding radii to $11.65$ and $11.32$~km, and $R_{1.4}$ to $13.02$ and $12.64$~km. Across the range we study, $M_{\max}$ decreases almost linearly, by about $0.024\,\Msun$ per percent of DM, and $R_{1.4}$ by about $0.08$~km per percent. The whole $M$--$R$ curve moves toward smaller masses and radii, as earlier studies of DM cores also found~\cite{sandin2009dark,ciarcelluti2011,Ivanytskyi2020,das2022dark}. The reason is simple. The DM adds gravitating energy at the centre but does not share the pressure support of the nuclear matter around it, so a star of a given mass becomes more compact. Extending the sequences beyond $10\%$ gives $M_{\max}=2.080$, $2.049$ and $2.020\,\Msun$ at $\FDM=12$, $14$ and $16\%$, so $M_{\max}\geq2.01\,\Msun$ holds up to $\FDM\approx16\%$ and the range studied here lies well inside it.

In Fig.~\ref{fig:MR}(a) every sequence rises with $\varepsilon_c$ up to a clear maximum. A DM core lowers this maximum and moves it to higher central density, from $\varepsilon_c/c^2=1.95\times10^{15}$ to $2.66\times10^{15}\,\gcm$ between $\FDM=0$ and $10\%$. At low central density the curves differ mostly by a shift in $\varepsilon_c$. Part of this shift is simply bookkeeping, because $\varepsilon_c$ includes the DM, which provides $3$--$24\%$ of the central energy density at $M_{\max}$ (Table~\ref{tab:mmax}).

All sequences with $\FDM\leq10\%$ stay above the $1\sigma$ masses of PSR~J0348+0432 ($2.01\pm0.04\,\Msun$) and PSR~J0740+6620 ($2.08\pm0.07\,\Msun$)~\cite{antoniadis2013,Fonseca2021} and are compatible with PSR~J0952$-$0607 ($2.35\pm0.17\,\Msun$)~\cite{romani2022}; see also Fig.~\ref{fig:trends}(a). For PSR~J0740+6620, the radius we predict for a $2.08\,\Msun$ star decreases from $13.11$~km ($\FDM=0$) to $11.78$~km ($10\%$), and all of these values lie inside the 68\% interval $12.39^{+1.30}_{-0.98}$~km of Ref.~\cite{Riley2021}. The curves in Fig.~\ref{fig:MR}(b) also cross the NICER credible regions of PSR~J0030+0451~\cite{riley2019,miller2019nicer}. The tighter constraints come from the two measurements that prefer smaller radii. PSR~J0437$-$4715 has $R=11.36^{+0.95}_{-0.63}$~km at $1.42\,\Msun$~\cite{Choudhury:2024xbk}; all our sequences lie outside its 68\% interval, and the $10\%$ curve comes close to its 90\% boundary. The EoS-insensitive GW170817 radii~\cite{Abbott2018} are reached only by the $\FDM=10\%$ sequence, at the edge of their 90\% region. In other words, a DM core pushes the stiff HPU-3 star toward the radii favoured by these two measurements without breaking the maximum-mass limits. This is a direct example of how DM content and nuclear physics can imitate each other.

More recent NICER analyses point in different directions. A reanalysis of PSR~J0437$-$4715 that includes the modulated nonthermal emission gives a median radius of $14.1$~km with a 68\% range of $11.9$--$15.5$~km~\cite{Miller2026a}, which contains all our sequences [Fig.~\ref{fig:MR}(b)]. For PSR~J0614$-$3329, with $M\approx1.44\,\Msun$, the headline result $R=10.29^{+1.01}_{-0.86}$~km~\cite{Mauviard2025} lies well below all our models, which give $12.6$--$13.4$~km at that mass [Fig.~\ref{fig:MR}(b)]. An independent analysis of the same pulsar gives $9.88$--$12.77$~km~\cite{Miller2026b}, which only the $10\%$ sequence reaches. For PSR~J1231$-$1411 the inferred radius depends strongly on the radius prior~\cite{Salmi2024}. If radii near $10$~km at $1.4\,\Msun$ are confirmed, they would disfavour HPU-3 even with a $10\%$ DM core. A DM core of this size also cannot explain the light compact object in HESS~J1731$-$347, with $M=0.77^{+0.20}_{-0.17}\,\Msun$ and $R=10.4^{+0.86}_{-0.78}$~km~\cite{Doroshenko2022}: at $0.77\,\Msun$ our radii range from $13.48$~km without DM to $12.76$~km at $\FDM=10\%$.

\subsection{Internal structure and the dark-matter core}\label{sec:core}

\begin{figure*}[t]
\centering
\includegraphics[width=\textwidth]{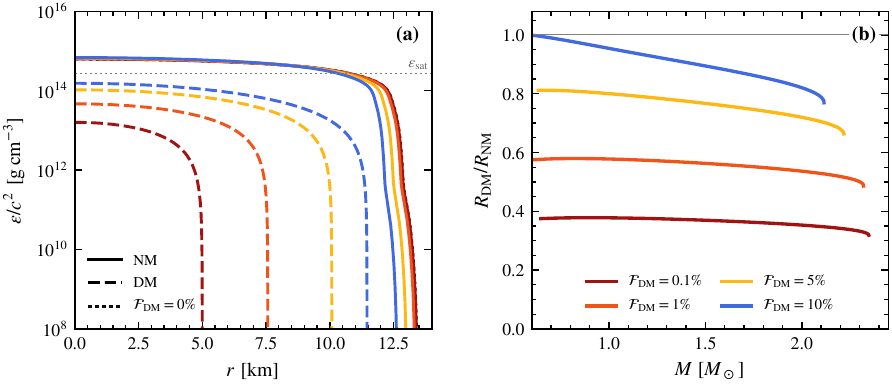}
\caption{(a) Energy-density profiles of nuclear matter (solid) and DM (dashed) in $1.4\,\Msun$ stars; colours give $\FDM$ as listed in panel (b), and the dotted horizontal line marks nuclear saturation, $\varepsilon_{\rm sat}/c^2=2.7\times10^{14}\,\gcm$. (b) Ratio of the DM-core radius to the nuclear-matter radius along the stable sequences. All configurations shown have $R_{\DM}<R_{\NM}$.}
\label{fig:structure}
\end{figure*}

Figure~\ref{fig:structure}(a) shows the radial profiles of the two fluids in $1.4\,\Msun$ stars. The DM energy density peaks at the centre, at $\approx1.6\times10^{13}\,\gcm$ for $\FDM=0.1\%$ and $\approx1.6\times10^{14}\,\gcm$ for $10\%$, and drops to zero at $R_{\DM}=5.0$, $7.6$, $10.1$ and $11.5$~km for $\FDM=0.1$, $1$, $5$ and $10\%$. In every case this is well inside the nuclear-matter surface. The nuclear matter near the centre is only mildly affected; the bigger change is in the outer layers, which are pulled inward as $\FDM$ grows.

Figure~\ref{fig:structure}(b) shows the size of the core through the ratio $R_{\DM}/R_{\NM}$. For $0.6\,\Msun\leq M\leq M_{\max}$ the ratio lies between $0.32$ and $0.38$ for $\FDM=0.1\%$, $0.49$ and $0.58$ for $1\%$, $0.66$ and $0.81$ for $5\%$, and $0.77$ and $1.00$ for $10\%$. It falls with mass because heavier stars squeeze the DM more. Every configuration with $M\geq0.6\,\Msun$ has $R_{\DM}<R_{\NM}$, so the DM always forms a core. Only for $\FDM=10\%$ and $0.41\lesssim M/\Msun\lesssim0.56$ does $R_{\DM}$ slightly exceed $R_{\NM}$, because there $h_{\DM}(0)$ comes close to $h_{\NM}(0)$. These light stars are outside the range of observed NS masses, and we leave them out of the figures. Since the DM never extends beyond the nuclear-matter surface, the increase of $\Lambda$ and $M$ found for extended DM distributions~\cite{Nelson2019,karkevandi2022,Giangrandi2023} does not appear here.

\subsection{Compactness and Love number}\label{sec:compactness}

\begin{figure*}[t]
\centering
\includegraphics[width=\textwidth]{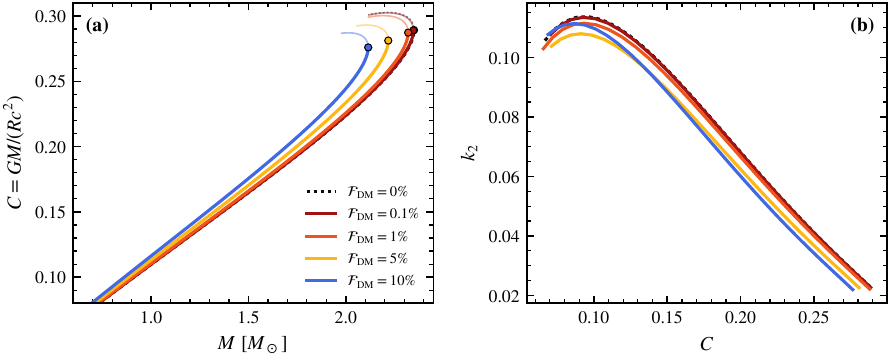}
\caption{(a) Compactness $C=GM/(Rc^2)$ as a function of mass; faded lines show the branches beyond the maximum mass and circles the maximum-mass configurations. (b) Tidal Love number $k_2$ as a function of compactness for stable configurations with $M\geq0.6\,\Msun$.}
\label{fig:Ck2}
\end{figure*}

At fixed mass a DM core raises the compactness [Fig.~\ref{fig:Ck2}(a)]: $C_{1.4}$ goes from $0.1538$ to $0.1635$ between $\FDM=0$ and $10\%$. Below about $1\,\Msun$ the curves nearly overlap, and the differences grow toward the maximum mass. If we compare the maximum-mass stars themselves, however, the compactness goes down with $\FDM$, from $0.2893$ to $0.2760$, because the maximum mass drops faster than its radius. The two trends do not contradict each other. The DM core makes a star of a given mass smaller, but it also removes the heaviest and most compact stars from the stable branch.

Along every sequence the Love number falls as the compactness grows [Fig.~\ref{fig:Ck2}(b)]. At fixed $C$ a DM core lowers $k_2$, because more of the mass sits near the centre and a centrally concentrated star responds less to a tidal field. At $1.4\,\Msun$, $k_2$ drops from $0.0939$ to $0.0808$. At the maximum mass, $k_2\simeq0.0222$--$0.0225$ hardly depends on $\FDM$, since in Eq.~\eqref{eq:k2} the increase of $y_R$ and the decrease of $C$ cancel each other.

\subsection{Tidal deformability}\label{sec:tidal_results}

\begin{figure}[t]
\centering
\includegraphics[width=\columnwidth]{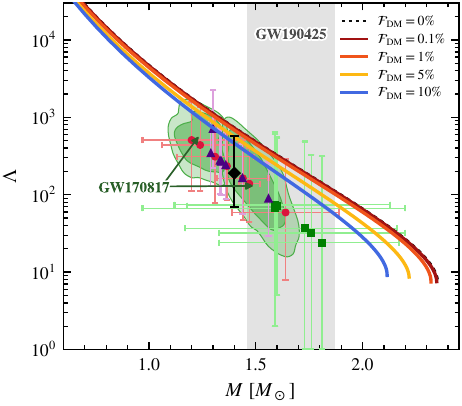}
\caption{Dimensionless tidal deformability as a function of mass. The green regions are the 68\% and 90\% credible regions for the two components of GW170817 (EoS-insensitive analysis~\cite{Abbott2018}), and the black diamond is $\Lambda_{1.4}=190^{+390}_{-120}$ (90\%) from the same analysis. The grey band is the 90\% component-mass range of GW190425 for low spins, $1.46$--$1.87\,\Msun$~\cite{Abbott2020GW190425}. Circles, squares and triangles are $\Lambda$ values inferred for millisecond pulsars (MSP), low-mass X-ray binaries (LMXB) and double neutron stars (DNS) with universal relations calibrated on GW170817~\cite{KumarLandry2019}.}
\label{fig:Lambda}
\end{figure}

\begin{figure*}[t]
\centering
\includegraphics[width=\textwidth]{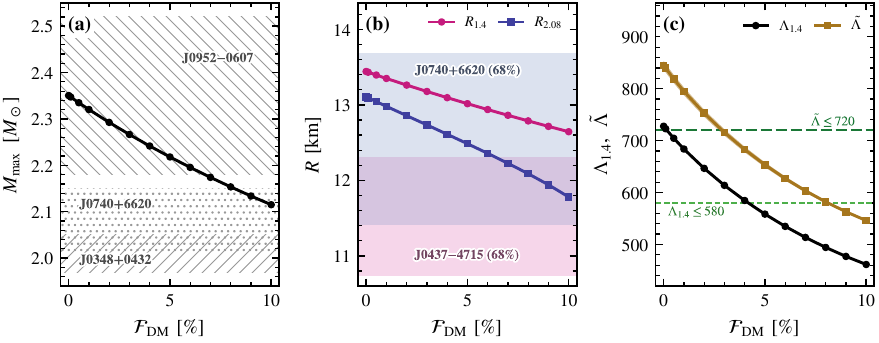}
\caption{Dependence of key observables on the DM fraction (13 sequences between $0$ and $10\%$). (a) Maximum mass, with the $1\sigma$ masses of PSR~J0348+0432~\cite{antoniadis2013}, PSR~J0740+6620~\cite{Fonseca2021} and PSR~J0952$-$0607~\cite{romani2022}. (b) Radii of $1.4$ and $2.08\,\Msun$ stars, with the 68\% radius intervals of PSR~J0437$-$4715~\cite{Choudhury:2024xbk} and PSR~J0740+6620~\cite{Riley2021}. (c) $\Lambda_{1.4}$ and the binary tidal deformability $\tilde\Lambda$ of GW170817 (band: $0.73\leq q\leq1$), with the 90\% upper bounds $\Lambda_{1.4}\leq580$~\cite{Abbott2018} and $\tilde\Lambda\leq720$~\cite{Abbott2019PRX}.}
\label{fig:trends}
\end{figure*}

Figure~\ref{fig:Lambda} shows $\Lambda(M)$ together with the observational constraints. For all DM fractions $\Lambda$ falls steeply with mass, by almost four orders of magnitude between $0.6\,\Msun$ and $M_{\max}$. A DM core lowers $\Lambda$ at every mass, since $\Lambda\propto k_2C^{-5}$ and the core both raises $C$ and lowers $k_2$. The effect grows with mass. For $\FDM=5$--$10\%$ the reduction is $20$--$30\%$ at $1\,\Msun$, $23$--$37\%$ at $1.4\,\Msun$, and more than $60\%$ near the maximum mass [Fig.~\ref{fig:shifts}(b)]. For $\FDM\leq1\%$ the change at $1.4\,\Msun$ is at most $6\%$.

At the maximum mass the trend goes the other way. $C(M_{\max})$ decreases with $\FDM$ while $k_2(M_{\max})$ barely moves, so $\Lambda(M_{\max})$ rises from $7.40$ to $9.23$ (Table~\ref{tab:mmax}). This only reflects the fact that the sequence now ends at a lower mass; it does not mean that DM makes a star of a given mass easier to deform.

The comparison with GW170817 is the most interesting result of this section. Without DM, HPU-3 gives $\Lambda_{1.4}=728$, above the 90\% upper bound $\Lambda_{1.4}\leq580$ from the EoS-insensitive analysis~\cite{Abbott2018}. It does satisfy the earlier bound $\Lambda(1.4\,\Msun)\leq800$ from the first GW170817 analysis~\cite{abbott2017gw170817}, which is the bound used in Ref.~\cite{kumar2023observational}. For the measured chirp mass and range of mass ratios it gives $\tilde\Lambda=837$--$853$, also above the bound $\tilde\Lambda=300^{+420}_{-230}$ (90\%, low-spin prior)~\cite{Abbott2019PRX}. In Fig.~\ref{fig:Lambda} the nucleonic curve runs along the upper edge of the GW170817 regions. A DM core removes this tension [Fig.~\ref{fig:trends}(c)]: $\tilde\Lambda$ drops below $720$ once $\FDM\gtrsim3.0\%$, and $\Lambda_{1.4}$ below $580$ once $\FDM\gtrsim4.2\%$. At these fractions $M_{\max}\approx2.24$--$2.27\,\Msun$, which still meets the radio-timing mass limits [Fig.~\ref{fig:trends}(a)]. We should be clear that this is a comparison with credible intervals and not a statistical inference. We also do not read it as evidence for DM. A nuclear EoS that fails the tidal test on its own can be brought within the bounds by adding DM, so the result says more about the degeneracy between DM content and the nuclear EoS than about the presence of DM~\cite{Kalita2026}. What it does show is that tidal data alone cannot distinguish a stiff nucleonic EoS with a few percent of DM in its core from a softer EoS without DM. This agrees with the conclusion of Ref.~\cite{Rutherford2025} that DM cores and the uncertainty in the nuclear EoS are hard to separate.

GW190425 gives only a weak constraint on $\Lambda$. For low spins its component masses lie between $1.46$ and $1.87\,\Msun$~\cite{Abbott2020GW190425}, where our sequences give $\Lambda\approx60$--$570$. The pulsar values obtained through universal relations~\cite{KumarLandry2019} have large error bars and cannot distinguish between the DM fractions we consider.

\subsection{Moment of inertia}\label{sec:inertia}

\begin{figure*}[t]
\centering
\includegraphics[width=\textwidth]{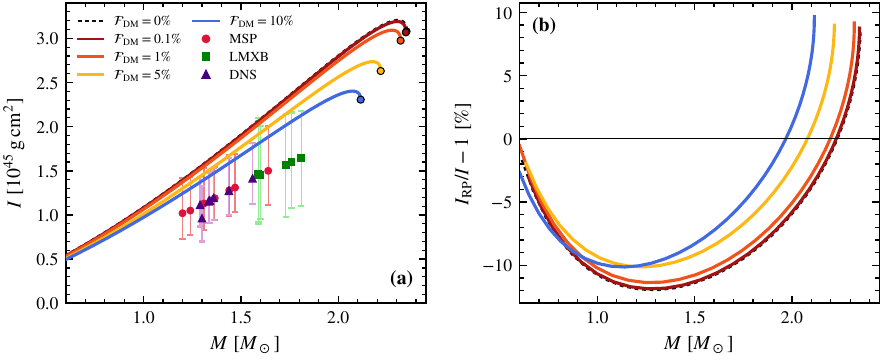}
\caption{(a) Moment of inertia from the slow-rotation equations [Eqs.~\eqref{eq:z-ode}--\eqref{eq:I}] versus mass; the large circles at the ends of the curves mark the maximum-mass configurations. Symbols with error bars are values inferred from GW170817 with universal relations for observed millisecond pulsars (MSP), low-mass X-ray binaries (LMXB) and double neutron stars (DNS)~\cite{KumarLandry2019}, with the same symbols as in Fig.~\ref{fig:Lambda}. (b) Relative error of the Ravenhall--Pethick estimate, Eq.~\eqref{eq:I_RP}, compared with the full calculation.}
\label{fig:inertia}
\end{figure*}

We compute the moment of inertia in Hartle's slow-rotation approximation~\cite{Hartle1967}. For a star rotating uniformly with angular velocity $\Omega$, the frame-dragging function $\bar\omega=\Omega-\omega$ obeys
\begin{equation}
\frac{1}{r^{4}}\frac{d}{dr}\left(r^{4}j\frac{d\bar\omega}{dr}\right)+\frac{4}{r}\frac{dj}{dr}\,\bar\omega=0,\qquad
j=e^{-\nu}\sqrt{1-2m/r},
\label{eq:hartle}
\end{equation}
with $dj/dr=-4\pi r\,e^{\lambda}(\varepsilon+P)\,j$, where $\varepsilon$ and $P$ are the totals of both fluids. We assume that the two fluids rotate together. Since they interact only through gravity, they could in principle rotate at different rates~\cite{KumarSotani2026}. The DM carries only a small share of $\int\varepsilon r^{4}dr$, namely $0.3\%$, $2.8\%$ and $7.5\%$ in $1.4\,\Msun$ stars with $\FDM=1$, $5$ and $10\%$, so a non-rotating DM core would lower $I$ by roughly these amounts. Introducing $z=r\bar\omega'/\bar\omega$ turns Eq.~\eqref{eq:hartle} into the first-order equation
\begin{equation}
\frac{dz}{dr}=-\frac{z(z+3)}{r}+4\pi r\,e^{\lambda}(\varepsilon+P)(z+4),\qquad z(0)=0,
\label{eq:z-ode}
\end{equation}
which we integrate together with Eqs.~\eqref{eq:tov2-m}--\eqref{eq:tov2-p} and \eqref{eq:y-ode}. Outside the star $\bar\omega=\Omega-2J/r^{3}$, and matching gives the angular momentum $J=R^{4}\bar\omega'(R)/6$ and
\begin{equation}
I=\frac{J}{\Omega}=\frac{R^{3}z_R}{6+2z_R}.
\label{eq:I}
\end{equation}
For comparison we also use the empirical estimate of Ravenhall and Pethick~\cite{RavenhallPethick1994,Lattimer2001},
\begin{equation}
I_{\rm RP}\simeq\frac{0.21\,MR^{2}}{1-2GM/(Rc^{2})},
\label{eq:I_RP}
\end{equation}
which needs only $M$ and $R$.

Figure~\ref{fig:inertia}(a) shows that $I$ grows with mass along each stable sequence and decreases with $\FDM$ at fixed mass. For $1.4\,\Msun$ stars, $I$ goes from $1.729\times10^{45}$ to $1.700$, $1.608$ and $1.529\times10^{45}\,\mathrm{g\,cm^2}$ for $\FDM=1$, $5$ and $10\%$, a drop of $1.7\%$, $7.0\%$ and $11.6\%$. At the maximum mass the drop is larger, from $3.079$ to $2.306\times10^{45}\,\mathrm{g\,cm^2}$ ($-25\%$), because both $M_{\max}$ and $R(M_{\max})$ go down and $I\sim MR^{2}$. Mass in the DM core sits close to the centre, where it adds little to $I$, so moving mass from the envelope into the core lowers $I$ more than it lowers $M$.

Figure~\ref{fig:inertia}(b) shows that Eq.~\eqref{eq:I_RP} is good only to about $\pm10\%$ for these stars. It underestimates $I$ by up to $10$--$12\%$ at $M\approx1.1$--$1.3\,\Msun$ and overestimates it by up to $10\%$ near $M_{\max}$, with the sign change between $1.96$ and $2.21\,\Msun$ depending on $\FDM$. Its accuracy also changes with $\FDM$, because a DM core reshapes the density profile in a way that $M$ and $R$ alone cannot capture. The changes caused by the DM are of similar size, $12$--$25\%$ for $\FDM=10\%$, so the full calculation is needed. A future measurement of the moment of inertia of PSR~J0737$-$3039A ($M=1.338\,\Msun$) from double-pulsar timing~\cite{kramer2021strong}, with $\sim5\%$ precision, would be sensitive to DM fractions of a few percent, provided the nuclear EoS were known independently.

In the dimensionless form $\bar I=I/M^3$, the I--Love relation of Yagi and Yunes~\cite{Yagi2013,Yagi2013Science} stays universal for all $\FDM$ we consider when the two fluids co-rotate [Fig.~\ref{fig:universal}(a)]. The deviations from the fit remain within $1\%$ for $\Lambda<10^4$, and the curves for different DM fractions differ from each other by less than $0.6\%$. The compactness--Love relation~\cite{Maselli2013} holds up less well [Fig.~\ref{fig:universal}(b)]. A DM core shifts $C(\Lambda)$ by up to $1.5\%$ ($\FDM=5\%$) and $2.5\%$ ($10\%$) relative to the nucleonic sequence, which is about as large as the EoS spread of that relation. If the fluids rotated independently and only the rotation of ordinary matter were observable, the I--Love--Q relations could deviate much more~\cite{Cronin2023}.

\subsection{Overall impact of the DM fraction}\label{sec:trends}

\begin{figure*}[t]
\centering
\includegraphics[width=\textwidth]{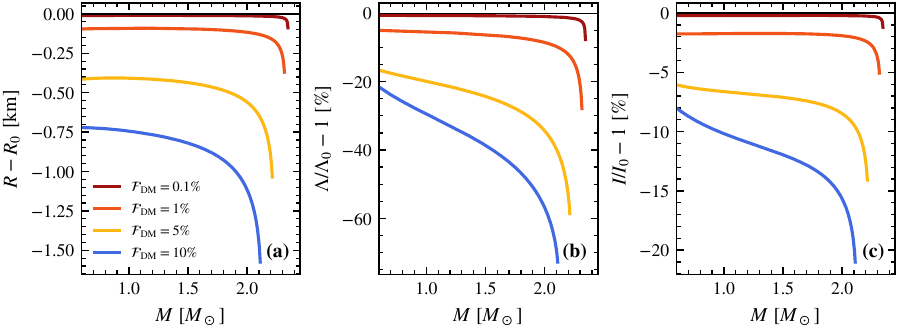}
\caption{Changes relative to the purely nucleonic star of the same mass (subscript 0): (a) radius, (b) tidal deformability and (c) moment of inertia, for $\FDM=0.1$, $1$, $5$ and $10\%$.}
\label{fig:shifts}
\end{figure*}

Figure~\ref{fig:shifts} compares stars of the same mass with and without DM, which is the comparison that matters for observations. Below $1.8\,\Msun$ the radius shrinks almost uniformly, by $0.01$, $0.1$, $0.4$ and $0.7$~km for $\FDM=0.1$, $1$, $5$ and $10\%$, and more strongly close to the maximum mass. The tidal deformability reacts much more than the radius: for $\FDM=5\%$ the radius of a $1.4\,\Msun$ star changes by $3\%$, while $\Lambda_{1.4}$ changes by $23\%$. The moment of inertia falls in between, at $7\%$. For $\FDM\leq0.1\%$ every change is below $1\%$, far below current observational uncertainties.

\subsection{Dependence on the DM parameters}\label{sec:sensitivity}

\begin{table}[t]
\caption{Sensitivity to the DM parameters at $\FDM=5\%$, with the Walecka EoS evaluated directly for each set. The first row is the baseline of Table~\ref{tab:dm_couplings}; the next six vary one parameter over its $1\sigma$ posterior range from Ref.~\cite{das2022dark}. $M_D$ is in MeV and the coupling ratios in GeV$^{-1}$. $\tilde\Lambda$ is for GW170817. The last column is $R_{\DM}/R_{\NM}$ of the $1.4\,\Msun$ star; values above 1 mean a DM halo.}
\label{tab:sensitivity}
\begin{ruledtabular}
\begin{tabular}{lcccc}
DM parameters & $M_{\max}/\Msun$ & $\Lambda_{1.4}$ & $\tilde\Lambda$ & $R_{\DM}/R_{\NM}$\\
\hline
Baseline & 2.218 & 559 & 649--659 & 0.77\\
$M_D=626.6$ & 2.225 & 572 & 665--674 & 0.85\\
$M_D=863.4$ & 2.212 & 548 & 636--646 & 0.71\\
$g_{sD}/m_{sD}=3.20$ & 2.219 & 559 & 650--660 & 0.78\\
$g_{sD}/m_{sD}=4.72$ & 2.218 & 558 & 648--657 & 0.77\\
$g_{vD}/m_{vD}=11.42$ & 2.217 & 557 & 647--657 & 0.77\\
$g_{vD}/m_{vD}=12.41$ & 2.220 & 561 & 652--661 & 0.79\\
No self-interaction & 2.194 & 531 & 615--625 & 0.60\\
$M_D=2000$ & 2.191 & 520 & 602--612 & 0.45\\
$M_D=1200$ & 2.202 & 533 & 618--628 & 0.59\\
$M_D=600$ & 2.227 & 576 & 670--679 & 0.87\\
$M_D=500$ & 2.237 & 599 & 699--707 & 0.99\\
$M_D=400$ & 2.252 & 647 & 760--765 & 1.23\\
\end{tabular}
\end{ruledtabular}
\end{table}

To see how much the results depend on the DM parameters, we recomputed the $\FDM=5\%$ sequence with the Walecka EoS of Sec.~\ref{sec:fdm} evaluated directly for each parameter set (Table~\ref{tab:sensitivity}). With the baseline parameters this reproduces the results of the analytic fit. Varying $M_D$, $g_{sD}/m_{sD}$ or $g_{vD}/m_{vD}$ over its $1\sigma$ posterior range changes $M_{\max}$ by at most $0.007\,\Msun$, $\Lambda_{1.4}$ by at most $13$ ($2\%$) and $R_{1.4}$ by at most $0.02$~km. The GW170817 thresholds of Sec.~\ref{sec:tidal_results} therefore move by only a few tenths of a percent in $\FDM$. Switching the self-interaction off lowers $\Lambda_{1.4}$ from $559$ to $531$.

The particle mass matters much more than the couplings. At $\FDM=5\%$ every star on the stable branch has a DM core for $M_D\gtrsim0.6$~GeV. Below $M_D\approx0.5$~GeV the $1.4\,\Msun$ star carries a DM halo, $\Lambda$ increases instead of decreasing, and $\tilde\Lambda$ rises above the GW170817 bound ($760$--$765$ at $M_D=400$~MeV). This agrees with the systematic study of Ref.~\cite{Karan2026}, in which light DM forms halos and heavy DM forms cores. Our baseline mass lies about $2\sigma$ above this transition, so the core picture of this paper is robust within the posterior of Ref.~\cite{das2022dark}, but it does not extend to sub-GeV DM in general.

\subsection{A softer nuclear equation of state}\label{sec:sly}

\begin{table}[t]
\caption{DM-admixed stars with the SLy EoS~\cite{DouchinHaensel2001}, using the analytic fit of Ref.~\cite{HaenselPotekhin2004} and the baseline DM parameters. Masses are in $\Msun$ and radii in km; $\tilde\Lambda$ is for GW170817.}
\label{tab:sly}
\begin{ruledtabular}
\begin{tabular}{cccccc}
$\FDM$ & $M_{\max}$ & $R(M_{\max})$ & $R_{1.4}$ & $\Lambda_{1.4}$ & $\tilde\Lambda$\\
\hline
$0$     & 2.048 & 9.98 & 11.70 & 297 & 356\\
$1\%$   & 2.025 & 9.92 & 11.61 & 277 & 332--333\\
$2\%$   & 2.003 & 9.86 & 11.53 & 260 & 312--313\\
$3\%$   & 1.982 & 9.80 & 11.45 & 245 & 295--297\\
$5\%$   & 1.943 & 9.69 & 11.29 & 220 & 266--270\\
\end{tabular}
\end{ruledtabular}
\end{table}

To see which conclusions depend on the stiffness of HPU-3, we repeated the calculation with the SLy EoS~\cite{DouchinHaensel2001}, a unified EoS that describes the crust and the core with one Skyrme interaction. We use the analytic fit of Ref.~\cite{HaenselPotekhin2004}, built on the same idea as our fits in Sec.~\ref{sec:fit}. Without DM our solver gives $M_{\max}=2.048\,\Msun$, in agreement with the value $2.05\,\Msun$ of Ref.~\cite{DouchinHaensel2001}, together with $R_{1.4}=11.70$~km and $\Lambda_{1.4}=297$ (Table~\ref{tab:sly}). Unlike HPU-3, SLy satisfies the GW170817 tidal bounds on its own ($\tilde\Lambda=356$).

The trends with $\FDM$ are the same as for HPU-3. The DM forms a core ($R_{\DM}/R_{\NM}=0.58$ at $1.4\,\Msun$ for $\FDM=1\%$), and $R_{1.4}$, $\Lambda_{1.4}$ and $M_{\max}$ all decrease, the last by about $0.021\,\Msun$ per percent of DM. What changes is the room left by the heavy pulsars. Because $M_{\max}$ starts only $0.04\,\Msun$ above $2.01\,\Msun$, it drops below that value already at $\FDM\approx1.7\%$. For a soft EoS it is therefore the heavy pulsars, not the tidal data, that limit the DM fraction. This is the headroom argument of Ref.~\cite{Kalita2026} in concrete form: how much DM a neutron star can hold depends mainly on how far the maximum mass of the nuclear EoS lies above the observed heavy pulsars. Taken together, HPU-3 and SLy bracket the behaviour expected for realistic nucleonic EoSs.

\section{Conclusions}\label{sec:conclusions}

We have studied neutron stars that contain fermionic DM coupled to nuclear matter only through gravity. Nucleonic matter was described by the HPU-3 RMF EoS and DM by a Walecka-type EoS with scalar and vector self-interactions, and we computed the structure, tidal deformability and moment of inertia for DM mass fractions up to $10\%$. Our main findings are the following.

(i) In all stars with $M\geq0.6\,\Msun$ the DM stays in a core inside the nuclear-matter surface, with $R_{\DM}/R_{\NM}\approx0.3$--$1.0$ depending on $\FDM$ and $M$. The core forms because the DM starts with a smaller central pseudo-enthalpy than the nuclear matter, so it runs out first.

(ii) The DM core makes the star more compact. From $\FDM=0$ to $10\%$ the maximum mass falls from $2.351$ to $2.115\,\Msun$ and its radius from $12.00$ to $11.32$~km, while $R_{1.4}$ falls from $13.44$ to $12.64$~km. The drop in $M_{\max}$ is close to linear, about $0.024\,\Msun$ per percent of DM. Fractions $\FDM\lesssim0.1\%$ have no observable effect, and $\FDM=1\%$ changes the global properties by a few percent at most.

(iii) At fixed mass both the Love number and the tidal deformability decrease with $\FDM$; $\Lambda_{1.4}$ goes from $728$ to $461$. At the maximum mass $\Lambda$ rises slightly, from $7.4$ to $9.2$, but only because the sequence ends at a lower mass. Pure HPU-3 violates the GW170817 bounds on $\Lambda_{1.4}$ and $\tilde\Lambda$. A DM core with $\FDM\gtrsim3$--$4\%$ brings the model within these bounds while keeping $M_{\max}>2.2\,\Msun$, in line with the observed massive pulsars. We read this as a degeneracy between DM content and the nuclear EoS rather than as evidence for DM.

(iv) The moment of inertia from the slow-rotation equations drops by $12\%$ at $1.4\,\Msun$ and by $25\%$ at the maximum mass for $\FDM=10\%$. The widely used Ravenhall--Pethick estimate is accurate only to about $10\%$ for these stars, which is not enough to resolve the effect of DM. For co-rotating fluids the I--Love relation remains universal with a DM core, whereas the C--Love relation shifts by up to $2.5\%$.

(v) These results hold across the $1\sigma$ range of the DM parameters of Ref.~\cite{das2022dark}. A DM particle lighter than about $0.5$~GeV, however, forms a halo instead, which raises $\Lambda$.

(vi) With the softer SLy EoS the DM again forms a core and makes the star more compact, but $M_{\max}$ falls below $2.01\,\Msun$ at $\FDM\approx1.7\%$. How much DM is allowed is set mainly by the maximum-mass headroom of the nuclear EoS.

Taken together, these results show that a DM core and a softer nuclear EoS have very similar effects: a stiff nuclear EoS with a few percent of DM gives stars much like those of a softer EoS without DM. Combining tidal measurements, which react most strongly to DM, with radius measurements and a future moment-of-inertia measurement of PSR~J0737$-$3039A could help break this degeneracy.

Our study has some clear limitations. We used HPU-3 as the main nuclear EoS, with SLy only as a comparison, and, apart from the test of Sec.~\ref{sec:sensitivity}, one set of DM parameters. The DM couples to nuclear matter only through gravity, the stars are static or slowly rotating, and the two fluids were assumed to co-rotate. Stability was judged from the maximum mass along fixed-$\FDM$ sequences rather than from radial modes, and the comparison with observations used credible intervals rather than a statistical inference. The SLy comparison and two-fluid studies with many nuclear EoSs~\cite{Thakur2023} suggest that the qualitative trends found here are generic, while the thresholds are specific to the nuclear EoS. The newest NICER results enter only as 68\% intervals, because their full posteriors were not available to us. In future work we plan to explore a wider range of nuclear EoSs and DM particle masses, including the halo regime, and to carry out a Bayesian analysis of gravitational-wave and NICER data to constrain the allowed DM fraction.

\section*{Data availability}
The observational data used in this work are publicly available from
the corresponding mission archives.

\appendix

\section{Model parameters}\label{app:params}

The HPU-3 parametrization~\cite{kumar2023observational} was designed to support the mass of PSR~J0952$-$0607 while reproducing finite-nuclei properties and bulk nuclear-matter constraints. Its parameters are given in Table~\ref{tab:HPU3parameters}. As in the original paper, $\overline{\kappa}=\kappa/M_N$, $\overline{\lambda}=\lambda$, $a_1$, $a_2$, $b_1$ and $b_2$ are listed after multiplication by $10^{2}$. Table~\ref{tab:dm_couplings} gives the DM parameters; since the EoS depends only on the ratios of couplings to mediator masses, three numbers are enough. The coefficients of the analytic EoS fits, Eqs.~\eqref{eq:zeta_low}--\eqref{eq:zeta_total}, are listed in Table~\ref{tab:fitcoeffs}.

\begin{table}[h]
\caption{HPU-3 RMF parameter set~\cite{kumar2023observational}. The nucleon and meson masses are in MeV.}
\label{tab:HPU3parameters}
\begin{ruledtabular}
\begin{tabular}{lc}
Parameter & Value \\
\colrule
$M_N$ & 939.0 \\
$m_{\sigma}$ & 498.638 \\
$m_{\omega}$ & 782.5 \\
$m_{\rho}$ & 770.0 \\
$g_{\sigma}$ & 9.87733 \\
$g_{\omega}$ & 12.45229 \\
$g_{\rho}$ & 11.34072 \\
$\overline{\kappa}\ (\times10^{2})$ & 3.10148 \\
$\overline{\lambda}\ (\times10^{2})$ & $-1.57402$ \\
$\zeta$ & 0.003301 \\
$\Lambda_{v}$ & 0.00248 \\
$a_1\ (\times10^{2})$ [fm$^{-1}$] & 0.03559 \\
$a_2\ (\times10^{2})$ & 0.01428 \\
$b_1\ (\times10^{2})$ [fm$^{-1}$] & 0.47507 \\
$b_2\ (\times10^{2})$ & 0.92769 \\
\end{tabular}
\end{ruledtabular}
\end{table}

\begin{table}[h]
\caption{Parameters of the fermionic DM EoS, Eqs.~\eqref{eq:DM_energy_density}--\eqref{eq:DM_scalar_density}.}
\label{tab:dm_couplings}
\begin{ruledtabular}
\begin{tabular}{lcc}
Parameter & Symbol & Value \\
\colrule
DM fermion mass & $M_D$ & $730.57$ MeV \\
Scalar coupling ratio & $g_{sD}/m_{sD}$ & $3.90$ GeV$^{-1}$ \\
Vector coupling ratio & $g_{vD}/m_{vD}$ & $11.88$ GeV$^{-1}$ \\
\end{tabular}
\end{ruledtabular}
\end{table}

\begin{table*}[t]
\caption{Coefficients of the analytic EoS representation, Eqs.~\eqref{eq:zeta_low}--\eqref{eq:zeta_total}, for HPU-3 and for the fermionic DM EoS. The $c_i$ define the low-density part and the switching points $c_{11}$, $c_{12}$; the $\alpha_i$ define the high-density part and the switching widths $\alpha_1$, $\alpha_2$.}
\label{tab:fitcoeffs}
\begin{ruledtabular}
\begin{tabular}{crrcrr}
 & \multicolumn{1}{c}{HPU-3} & \multicolumn{1}{c}{Fermionic DM} & & \multicolumn{1}{c}{HPU-3} & \multicolumn{1}{c}{Fermionic DM}\\
\colrule
$c_1$    & 2.409894909275  & 4.30239422  & $\alpha_1$    & 4.609864060021  & 3.25810515 \\
$c_2$    & 7.486131398215  & 4.27809560  & $\alpha_2$    & 4.430346618589  & 3.21932286 \\
$c_3$    & $-0.617171119339$ & 0.10000230  & $\alpha_3$    & $-37.608320575890$ & 5.38164443 \\
$c_4$    & 0.517863186186  & 0.72170418  & $\alpha_4$    & 2.599338711648  & $-0.49186715$ \\
$c_5$    & 3.698578161798  & 1.62399695  & $\alpha_5$    & 6.976281760983  & 3.40865740 \\
$c_6$    & 11.897747836530 & 11.67018809 & $\alpha_6$    & 14.569689832450 & 14.24309874 \\
$c_7$    & 20.000000000000 & 10.24685636 & $\alpha_7$    & 5.581031227399  & 21.11963434 \\
$c_8$    & 0.855043904233  & 1.67691627  & $\alpha_8$    & 5.109350034203  & 1.04555187 \\
$c_9$    & 3.612834502674  & 1.61506047  & $\alpha_9$    & $-0.213529591027$ & 0.00830816 \\
$c_{10}$ & 11.903080262200 & 11.66881051 & $\alpha_{10}$ & 8.513865809808  & 0.56068844 \\
$c_{11}$ & 13.986998088230 & 14.13941517 & $\alpha_{11}$ & 12.813287158380 & 8.86881444 \\
$c_{12}$ & 14.013647588440 & 14.15642876 & & & \\
\end{tabular}
\end{ruledtabular}
\end{table*}

\begin{figure*}[t]
\centering
\includegraphics[width=\textwidth]{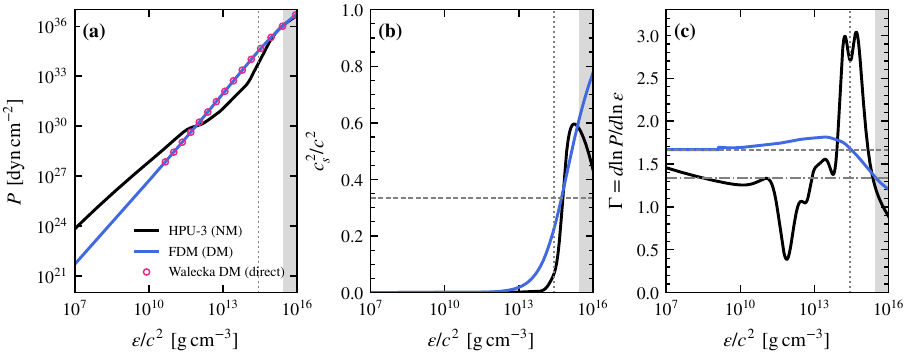}
\caption{(a) Pressure versus energy density for HPU-3 and the fermionic DM EoS; open circles are direct evaluations of the Walecka model with the parameters of Table~\ref{tab:dm_couplings}. (b) Squared sound speed; the dashed line marks the conformal value $1/3$. (c) Adiabatic index $\Gamma=d\ln P/d\ln\varepsilon$; the dashed and dash-dotted lines mark $5/3$ and $4/3$. The vertical dotted line indicates nuclear saturation and the grey band densities above the largest central density of a stable star.}
\label{fig:eos}
\end{figure*}

\begin{figure*}[t]
\centering
\includegraphics[width=\textwidth]{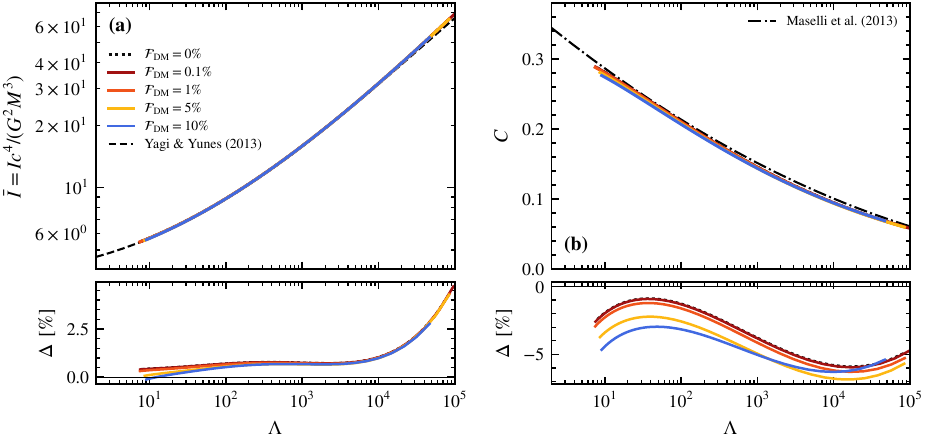}
\caption{(a) Normalized moment of inertia $\bar I=I/M^3$ versus $\Lambda$ with the fit of Yagi and Yunes~\cite{Yagi2013,Yagi2013Science}; the lower panel shows the relative deviation from the fit. (b) Compactness versus $\Lambda$ with the fit of Maselli et~al.~\cite{Maselli2013} and the relative deviation.}
\label{fig:universal}
\end{figure*}

\begin{figure*}[t]
\centering
\includegraphics[width=\textwidth]{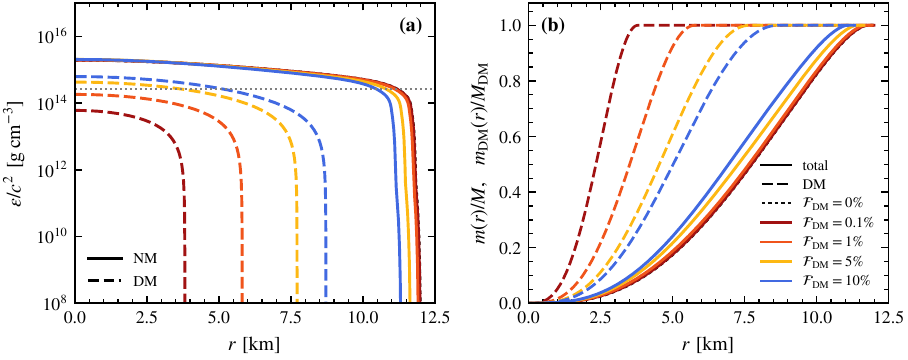}
\caption{Internal structure of the maximum-mass configurations. (a) Energy-density profiles of nuclear matter (solid) and DM (dashed), with colours as listed in panel (b). (b) Enclosed total mass $m(r)/M$ (solid) and enclosed DM mass $m_{\DM}(r)/M_{\DM}$ (dashed).}
\label{fig:profilesmax}
\end{figure*}

\begin{figure*}[t]
\centering
\includegraphics[width=\textwidth]{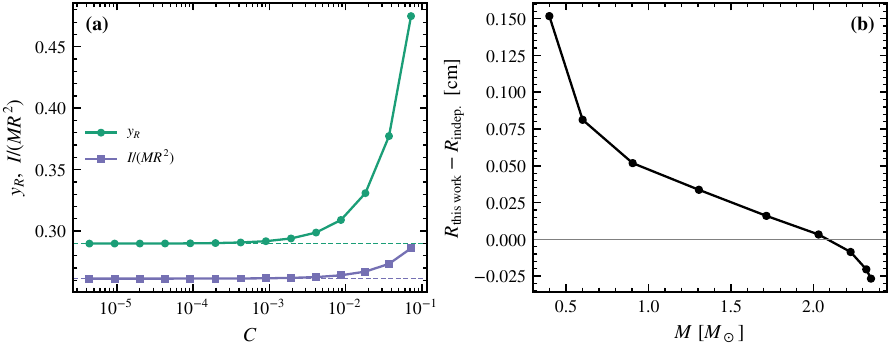}
\caption{Numerical tests. (a) $y_R$ and $I/(MR^2)$ for an $n=1$ polytrope as functions of compactness; the dashed lines are the Newtonian values $0.28986$ and $0.26138$. (b) Difference between the radii obtained in this work and those of an independent single-fluid integration in $r$ for HPU-3.}
\label{fig:validation}
\end{figure*}

\section{Numerical method and verification}\label{app:verification}

We integrate the equations as follows. For each fluid we tabulate $\varepsilon_i$, $P_i$ and $d\varepsilon_i/dh_i=(\varepsilon_i+P_i)/c_{s,i}^{2}$ on a uniform grid in $\ln h_i$, where $h_i=\int_0^{P_i}dP/(\varepsilon_i+P)$ is computed by quadrature from the analytic EoS. With $t=h_i(0)-h_i(r)$ as independent variable, $dr/dt=r(r-2m)/(m+4\pi r^3P)$. We multiply Eqs.~\eqref{eq:tov2-m}, \eqref{eq:y-ode} and \eqref{eq:z-ode} by $dr/dt$ and integrate them with the adaptive LSODA method (relative tolerance $10^{-10}$), starting from a series expansion near $r=0$. The integration is split at $t=\min_i h_i(0)$, where the inner fluid ends. The source term in Eq.~\eqref{eq:Q} is computed as $\sum_i d\varepsilon_i/dh_i$, so no numerical derivatives of the EoS are needed. Fixing $\FDM$ at a given $\varepsilon_c$ takes a one-dimensional Brent root search in $\ln[\varepsilon_{\DM}(0)/\varepsilon_{\NM}(0)]$, usually about eight structure integrations.

To check the equations we compared our results with cases where the answer is known. The results are collected in Table~\ref{tab:benchmarks} and Fig.~\ref{fig:validation}.
\begin{itemize}
\item For an $n=1$ polytrope in the Newtonian limit ($C\sim10^{-6}$) we find $y_R=0.28988$ and $I/(MR^2)=0.26138$. The analytic values are $0.28986$, which corresponds to $k_2=(15-\pi^2)/(2\pi^2)$, and $\frac{2}{3}(1-6/\pi^2)=0.26138$. This checks Eqs.~\eqref{eq:y-ode}, \eqref{eq:Q}, \eqref{eq:z-ode} and \eqref{eq:I}.
\item For the standard general-relativistic $\Gamma=2$, $K=100$ polytrope with central rest-mass density $1.28\times10^{-3}$ ($G=c=\Msun=1$) we find $M=1.4002$, $R=9.5856$ and $M_{\max}=1.6373$, in agreement with the reference values $1.400$, $9.586$ and $1.637$.
\item For HPU-3 without DM, a separate integration of Eqs.~\eqref{eq:tov-m}--\eqref{eq:tov-p} in $r$, with $P$ as the dependent variable, gives the same radii to better than $0.2$~cm for $0.4\leq M/\Msun\leq2.35$ [Fig.~\ref{fig:validation}(b)].
\item Evaluating the Walecka model, Eqs.~\eqref{eq:DM_energy_density}--\eqref{eq:DM_scalar_density}, with the parameters of Table~\ref{tab:dm_couplings} reproduces the analytic DM EoS to $|\Delta\zeta|\leq10^{-3}$ for $9.1\leq\xi\leq16.5$ [open circles in Fig.~\ref{fig:eos}(a)]. If we instead fit the three Walecka parameters freely to the analytic EoS, we get $M_D=730.55$~MeV, $g_{sD}/m_{sD}=3.897$~GeV$^{-1}$ and $g_{vD}/m_{vD}=11.878$~GeV$^{-1}$. Table~\ref{tab:dm_couplings} and the fit coefficients therefore describe the same EoS.
\item For pure HPU-3 we find $M_{\max}=2.351\,\Msun$, $R(M_{\max})=12.00$~km, $R_{1.4}=13.44$~km, $R_{2.08}=13.11$~km and $\Lambda_{1.4}=728$. Reference~\cite{kumar2023observational} reports $2.35\,\Msun$, $11.97$~km, $13.39$~km, $12.98$--$13.09$~km (HPU models) and $719$.
\end{itemize}

\begin{table}[h]
\caption{Verification of the numerical implementation against known results.}
\label{tab:benchmarks}
\begin{ruledtabular}
\begin{tabular}{lcc}
Test & This work & Reference \\
\colrule
$n=1$ Newtonian limit: $y_R$ & 0.28988 & 0.28986 \\
$n=1$ Newtonian limit: $I/(MR^2)$ & 0.26138 & 0.26138 \\
$\Gamma=2$, $K=100$: $M$ & 1.4002 & 1.400 \\
$\Gamma=2$, $K=100$: $R$ & 9.5856 & 9.586 \\
$\Gamma=2$, $K=100$: $M_{\max}$ & 1.6373 & 1.637 \\
HPU-3: $M_{\max}\,[\Msun]$ & 2.351 & 2.35~\cite{kumar2023observational} \\
HPU-3: $R_{1.4}$ [km] & 13.44 & 13.39~\cite{kumar2023observational} \\
HPU-3: $\Lambda_{1.4}$ & 728 & 719~\cite{kumar2023observational} \\
Walecka DM vs fit: $\max|\Delta\zeta|$ & $1\times10^{-3}$ & 0 \\
\end{tabular}
\end{ruledtabular}
\end{table}

\section{Supplementary figures}\label{app:figures}

Figure~\ref{fig:eos} shows the two EoSs together with their sound speeds and adiabatic indices. HPU-3 softens at neutron drip ($\varepsilon/c^2\approx4\times10^{11}\,\gcm$) and becomes much stiffer above saturation, with $c_s^2$ reaching $0.6$ at $\approx2\times10^{15}\,\gcm$. The DM EoS is stiffer than HPU-3 for $8\times10^{11}\lesssim\varepsilon/c^2\lesssim1.5\times10^{15}\,\gcm$. Its adiabatic index falls from $\approx1.8$ toward $4/3$ as the dark fermions become relativistic, and it stays causal. Figure~\ref{fig:universal} compares our sequences with the I--Love and C--Love relations, Fig.~\ref{fig:profilesmax} shows the interior of the maximum-mass stars, and Fig.~\ref{fig:validation} illustrates the numerical tests.

\clearpage
\bibliographystyle{apsrev4-2}
\bibliography{reference}

\end{document}